\documentclass[aps,draft,showpacs]{revtex4}

\begin{document}

\title{Multi-Mass Scale RG Improvement of the Coleman-Weinberg Model}
\author{Chungku Kim}
\affiliation{ Department of Physics, College of Natural Science, 
Keimyung University, Taegu, KOREA}
\date{ \today}

\begin{abstract}
We obtain the RG improvement of the effective potential for the Coleman-Weinberg
model by resumming the leading-logarithms which have three different mass scales. 
Then we investigate the effect of the multi-mass scale on the prediction of the
magnitude of the Higgs boson mass by considering the two-loop effective potential.

\end{abstract}
\maketitle

\section{ INTRODUCTION}

The Coleman-Weinberg (CW) model \cite{CW} is the massless scalar electrodynamics where
the scalar field does not have a tree level mass and the spontaneous
symmetry breaking occurs from the effective potential \cite{EP} which is the
radiative correction to the classical potential. The CW model and its extension
to the more realistic model have been studied extensively
due to its predictive power for the magnitude of the Higgs boson mass. The leading
logarithms of the effective 
potential can be resummed by using the renormalization group(RG) \cite{RG} known as RG
improvement and recently, the RG improvement of the effective potential of the CW model 
has been obtained by the optimal form  \cite{Pert} which incorporates
all possible logarithms of single mass scale to the effective potential that is accessible
via RG methods.

However, actually the effective potential of the CW model has three different mass
scales and in this paper, we will obtain the complete RG improvement of the leading- 
logarithms of the effective potential for the CW model by using the method of RG
 improvement in case of the multi-mass scale \cite{Bando}. Then we will investigate
the prediction of the Higgs boson mass by using the two-loop effective potential 
where the difference between the case of single mass scale and that of multi-mass  
scale appears for the first time.

\section{RG improvement of the CW model}

In this section, we will first obtain the RG improvement of the CW model by
using the method of the RG improvement in case of the multi-mass scale. The
classical Lagrangian of the CW model is given by 
\begin{equation}
L=\frac{1}{2}(\partial _{\mu }\phi _{1}-eA_{\mu }\phi _{2})^{2}+\frac{1}{2}%
(\partial _{\mu }\phi _{2}+eA_{\mu }\phi _{1})^{2}-\frac{\lambda }{24}(\phi
_{1}^{2}+\phi _{2}^{2}).
\end{equation}
In this paper, we will use the parameters $x$ and $y$ defined by

\begin{equation}
x\equiv \frac{e^{2}}{4\pi ^{2}}\ \qquad \text{and\qquad }y\equiv \frac{%
\lambda }{4\pi ^{2}}
\end{equation}

The effective potential of the CW model is
independent of the renormalization mass scale $\mu $ and hence satisfies the
renormalization group equation 

\begin{equation}
\lbrack \mu \frac{\partial }{\partial \mu }+\beta _{x}\frac{\partial }{%
\partial x}+\beta _{y}\frac{\partial }{\partial y}+\gamma \phi \frac{%
\partial }{\partial \phi }]V_{eff}=0
\end{equation}
where

\begin{equation}
\phi ^{2}=\phi _{1}^{2}+\phi _{2}^{2}
\end{equation}
and the RG functions $\beta $ and $\gamma $ are given by

\begin{equation}
\beta _{f}=\mu \frac{df}{d\mu }=\kappa \text{ }\beta _{f}^{(1)}+\kappa ^{2}%
\text{ }\beta _{f}^{(2)}+\cdot \cdot \cdot (f=x,y)
\end{equation}
and

\begin{equation}
\gamma =\frac{\mu }{\phi }\frac{d\phi }{d\mu }=\kappa \text{ }\gamma
_{f}^{(1)}+\kappa ^{2}\text{ }\gamma _{f}^{(2)}+\cdot \cdot \cdot
\end{equation}
with $\kappa =(4\pi ^{2})^{-1}.$

By using the method of characteristics\cite{char}, we can see that the
effective potential satisfies

\begin{equation}
V(x,y,\phi ,\mu )=V(x(t),y(t),\phi (t),\mu (t))  \label{char}
\end{equation}
where 
\begin{equation}
\mu (t)=\mu e^{t} 
\end{equation}
and $x(t),y(t)$ and $\phi (t)$ is the solution of the differential equation
such that

\begin{widetext}
\begin{equation}
\frac{df(t)}{dt}=\beta _{f}(x(t),y(t))\ =\kappa \text{ }\beta
_{f}^{(1)}(x(t),y(t))+\kappa ^{2}\text{ }\beta _{f}^{(2)}(x(t),y(t))+\cdot
\cdot \cdot (f=x,y)
\end{equation}
\end{widetext}
and

\begin{widetext}
\begin{equation}
\frac{d\phi (t)}{dt}=\gamma (x(t),y(t))\phi (t)=\kappa \text{ }\gamma
_{f}^{(1)}(x(t),y(t))+\kappa ^{2}\text{ }\gamma _{f}^{(2)}(x(t),y(t))+\cdot
\cdot \cdot
\end{equation}
\end{widetext}

with the initial conditions $x(0)=x,y(0)=y$ and $\phi (0)=\phi $.

Since the CW model is a O(2) scalar field theory coupled to the U(1) gauge
field, the effective potential of the CW model contains the following three different mass
scales. 

\begin{widetext}
\begin{equation}
L_{1}\equiv \log \left( \frac{2\pi ^{2}y\phi ^{2}}{\mu ^{2}}\right) ,
L_{2}\equiv \log \left( \frac{2\pi ^{2}y\phi ^{2}}{3\mu ^{2}}\right) 
\text{and }L_{3}\equiv \log \left( \frac{4\pi ^{2}x\phi ^{2}}{\mu ^{2}}%
\right) .
\end{equation}
\end{widetext}

The resummation of these three different leading-logarithms can be done by following
the similar steps to the case of two different mass scales\cite{Bando} as
follows. First let us write the effective potential of the CW model as

\begin{widetext}
\begin{equation}
V(x,y,\phi ,\mu )=\frac{\pi ^{2}}{6}y \phi ^{4}+\sum_{l=1}^{\infty
}\kappa ^{l}\phi ^{4}\sum_{n=0}^{l}\sum_{ {{p,q,r }\atop{{p+q+r=n}}} 
 } f^{(l,n,p,q,r)}(x,y)L_{1}^{p}L_{2}^{q}L_{3}^{r}  \label{LL}
\end{equation}
\end{widetext}

where $l$ is the loop order and the case $n=l$ ( $n=l-1$ ) corresponds to the
leading- ( next-to-leading )logarithms etc. 

By noting that one can choose arbitrary value for $t$ in Eqs.(7-10), let
us rescale the variables $\kappa $ , $t$ and the mass scales $L_{i}(i=1,2,3)$
at both sides of these equations as 
\begin{equation}
\kappa \rightarrow \hbar \kappa ,\qquad L_{i}\rightarrow \frac{L_{i}}{\hbar }%
\qquad \text{and }\qquad t\rightarrow \frac{t}{\hbar }
\end{equation}
and then substitute the effective potential given in Eq.(12) into 
Eq.(7). Since the leading logarithms of the effective potential given in L.H.S.
of Eq.(12) does not change under this rescaling, we can obtain the
resummation of the leading-logarithms ( $V_{LL}$ ) by taking the order $\hbar ^{0}$
terms of the R.H.S. of Eq.(7) with the effective potential given in Eq.(12) . 
Then we obtain 

\begin{widetext}
\begin{equation}
V_{LL}=\frac{\pi ^{2}}{6}y_{0}(t)\phi _{0}(t)^{4}+\sum_{l=1}^{\infty
}\kappa ^{l}\phi _{0}(t)^{4} \sum_{{p,q,r} \atop{p+q+r=l%
}} f^{(l,l,p,q,r)}(x_{0}(t),y_{0}(t)) \\
\times (L_{1}-2t)^{p}(L_{2}-2t)^{q}(L_{3}-2t)^{r}  \nonumber
\end{equation}
\end{widetext}

Here the quantities with the subscript $0$ are the order $\hbar ^{0}$
solutions of the Eq.(9) and Eq.(10) under the rescaling of Eq.(13) so that

\begin{equation}
\frac{df_{0}(t)}{dt}=\kappa \text{ }\beta _{f}^{(1)}(x_{0}(t),y_{0}(t))\
(f=x,y)
\end{equation}
and

\begin{equation}
\frac{d\phi _{0}(t)}{dt}=\kappa \gamma ^{(1)}(x_{(0)}(t),y_{(0)}(t))\phi
_{(0)}(t)
\end{equation}
Since we can choose arbitrary value for the variable $t$, we choose 
\begin{equation}
t=\frac{L_{3}}{2}
\end{equation}
Then only those terms with $r=0$ in Eq.(14) survives and we obtain 

\begin{widetext}
\begin{equation}
V_{LL} =\frac{\pi ^{2}}{6}y_{0}(\frac{L_{3}}{2})\phi _{0}(\frac{L_{3}}{2}%
)^{4}+\sum_{l=1}^{\infty }\kappa ^{l}\phi _{0}(\frac{L_{3}}{2}%
)^{4}\sum_{ {{p,q}\atop{p+q=l}}  } 
f^{(l,l,p,q,0)}(x_{0}(\frac{L_{3}}{2}),y_{0}(\frac{L_{3}}{2}%
))(L_{1}-L_{3})^{p}(L_{2}-L_{3})^{q} 
\end{equation}
\end{widetext}
Since the contributions from the coupling with gauge fields ( $r\neq 0$ term in Eq.(14))
has vanished, this is nothing but the resummation of the leading-logarithms of the
effective potential for the O(2) scalar field theory with coupling constants 
$f_{0}(\frac{L_{3}}{2})$ $(f=x,y)$, classical field $\phi _{0}(\frac{L_{3}}{2%
})$ and the mass scales $(L_{1}-L_{3})$ and $(L_{2}-L_{3})$. By using the
results given in \cite{Bando} we obtain 
\begin{equation}
V_{LL}=\frac{\pi ^{2}}{6}\text{ }\frac{y_{0}(\frac{L_{3}}{2})\phi _{0}(\frac{%
L_{3}}{2})^{4}}{1-\frac{3}{8}y_{0}(\frac{L_{3}}{2})(L_{1}-L_{3})-\frac{1}{24}%
y_{0}(\frac{L_{3}}{2})(L_{2}-L_{3})} 
\end{equation}
In order to obtain the coupling constants $f_{0}(\frac{L_{3}}{2})$ $(f=x,y)$%
and the classical field $\phi _{0}(\frac{L_{3}}{2})$, we use the one-loop RG
functions of the CW model \cite{CW} to obtain 
\begin{equation}
\frac{dx_{0}(t)}{dt}=\beta_{x}^{(1)}x_{0}(t)^{2}=\frac{1}{6}x_{0}(t)^{2}
\end{equation}

\begin{equation}
\frac{d\phi _{0}(t)}{dt}= \gamma
^{(1)}x_{0}(t)\phi _{0}(t)=\frac{3}{4}x_{0}(t)\phi _{0}(t)
\end{equation}
and

\begin{widetext}
\begin{equation}
\frac{dy_{0}(t)}{dt}= \beta _{y}^{(1)}y_{0}(t)^{2}+\beta _{yx}^{(1)}x_{0}(t)y_{0}(t)
+\beta_{yxx}^{(1)}x_{0}(t)^{2}=\frac{5}{6}y_{0}(t)^{2}-3x_{0}(t)y_{0}(t)+9x_{0}(t)^{2}
\end{equation}
\end{widetext}

Eqs.(20) and (21) can be solved easily and we can obtain 
\begin{equation}
x_{0}(t)=\frac{x}{1-\frac{1}{6}xt}
\end{equation}

\begin{equation}
\phi _{0}(t)=\frac{\phi }{(1-\frac{1}{6}xt)^{9/2}}
\end{equation}
In order to obtain the solution of the Eq.(22), we write $y_0 (t)$ as \cite{Kie}
\begin{equation}
y_{0}(t)=-\frac{F^{\prime }(t)}{\beta _{y}^{(1)}F(t)}
\end{equation}
where $F(t)$ is an auxiliary function.

By substituting this expression into Eq.(22), we obtain 
\begin{equation}
F^{\prime \prime }(t)-\beta _{yx}^{(1)}x(t)F^{\prime }(t)+\beta
_{y}^{(1)}\beta _{yxx}^{(1)}x(t)^{2}F(t)=0
\end{equation}
This is a Euler differential equation\cite{Math} and by changing the
variable from $t$ to $z$ as 
\begin{equation}
z=\frac{\beta _{y}^{(1)}}{\beta _{x}^{(1)}}\log (1-\beta _{x}^{(1)}xt)
\end{equation}
we obtain 
\begin{equation}
\beta _{y}^{(1)}\frac{d^{2}F(z)}{dz^{2}}+(\beta _{yx}^{(1)}-\beta _{y}^{(1)})%
\frac{dF(z)}{dz}+\beta _{yxx}^{(1)}F(z)=0
\end{equation}
By solving this equation and by using the initial condition for $y(t)$ as $%
y(0)=y$, we finally obtain

\begin{equation}
y_{0}(t)=x\frac{a(y-bx)G(t)^{a\delta -1}-b(y-ax)G(t)^{b\delta -1}}{%
(y-bx)G(t)^{a\delta }-(y-ax)G(t)^{b\delta }}
\end{equation}
where $\delta \equiv \beta _{y}^{(1)}/\beta _{x}^{(1)}$ and 
\begin{equation}
G(t)\equiv \frac{x}{x(t)}=1-\frac{1}{6}xt
\end{equation}
and a and b are the two roots of the equation 
\begin{equation}
\beta _{y}^{(1)}p^{2}+(\beta _{yx}^{(1)}-\beta _{x}^{(1)\ })p+\beta
_{yxx}^{(1)}=0
\end{equation}
This result agrees with the one given in Ref. \cite{Bando} obtained by some other method.
By using Eq.(20) and Eq.(22), can obtain 
\begin{equation}
\frac{dR(t)}{dt}=x(t)[\beta _{y}^{(1)}R(t)^{2}+(\beta _{yx}^{(1)}-\beta
_{x}^{(1)\ })R(t)+\beta _{yxx}^{(1)}]
\end{equation}
where $R(t)$ is the ratio between the two parameters $x$ and $y$
defined by

\begin{equation}
R(t)=\frac{y_{0}(t)}{x_{0}(t)}
\end{equation}
Then we can see that the two roots a and b of Eq.(31) becomes the fixed point%
\cite{Fixed} of the ratio $R(t)$.

By substituting the coefficients of the one-loop RG functions of the CW
model given in Eq(20) and Eq.(22), we obtain the two roots a and b of Eq.(31) as 
\begin{equation}
a,b=\frac{19}{10}\pm i\frac{\sqrt{719}}{10}
\end{equation}
By substituting the two roots a and b into Eq.(29), we obtain $y_0 (t)$ as 
\begin{equation}
y_{0}(t)=\frac{x}{G(t)}\frac{y+\frac{1}{\sqrt{719}}(19y-108x)\tan (\frac{%
\sqrt{719}}{2}\ln G(t))}{x-\frac{1}{\sqrt{719}}(19x-10y)\tan (\frac{\sqrt{719%
}}{2}\ln G(t))}
\end{equation}
By substituting Eq.(24) and (35) into Eq.(19) we can obtain the
resummation of the leading-logarithms terms of the effective potential for
the CW model. 
In order to compare with the perturbative expansion of the optimal RG improvement
given in \cite{Pert}
where single mass scale was considered, let us substitute the same value for three
different mass scale as $L_{1}=L_{2}=L_{3}=L$ into the resummation of the 
leading-logarithms given in Eq.(19) where 
\begin{equation}
L\equiv \log (\frac{\phi ^{2}}{\mu ^{2}})
\end{equation}
Then  we obtain 
\begin{widetext}
\begin{equation}
V_{LL}=\frac{\ \pi ^{2}x\phi ^{4}}{6\ G(\frac{L}{2})^{19}}\ \frac{y+\frac{1}{%
\sqrt{719}}(19y-108x)\tan (\frac{\sqrt{719}}{2}\ln G(\frac{L}{2}))}{x-\frac{1%
}{\sqrt{719}}(19x-10y)\tan (\frac{\sqrt{719}}{2}\ln G(\frac{L}{2}))}
\end{equation}
\end{widetext}

Now we should expand the functions
appearing in Eq.(37) as a power series in $x$. By using Eq.(30), the
expansion as a power series in $x$ of the $G(\frac{L}{2})^{19}$ in the denominator of
above equation is straightforward and $\frac{1}{\sqrt{719}}%
\tan (\frac{\sqrt{719}}{2}\ln G(\frac{L}{2}))$ can be expanded as a power
series in $x$ as 
\begin{widetext}
\begin{equation}
\frac{1}{\sqrt{719}}\tan (\frac{\sqrt{719}}{2}\ln G(\frac{L}{2}))=-\frac{1}{%
24}Lx-\frac{1}{576}L^{2}x^{2}-\frac{241\ }{13824}L^{3}x^{3}+O(x^{4})
\end{equation}
\end{widetext}
\ By substituting these results into Eq.(37) and writing the power series
expansion of $V_{LL}$ in $x$ as 
\begin{equation}
V_{LL}=\frac{\ \pi ^{2}\phi ^{4}}{6\ }\ [y\ S_{0}(yL)+x\ S_{1}(yL)+x^{2}L\
S_{2}(yL)+O(x^{3})]
\end{equation}
it is easy to check that the resulting coefficient functions $S_{i}(yL)$
(i=0,1,2) coincides with the results of Ref.\cite{Pert} exactly. 
\newline
Finally, let us consider the effect of the multi-mass scales on the
prediction of the magnitude of the Higgs mass. For simplicity, we will
consider up to two loop order where the difference between the case of single mass
scale and that of multi-mass scale appears 
for the first time. By expanding the RG improved effective potential for the CW model 
given in Eq.(19) up to two loop order where $\phi_0 (t)$ and $y_0 (t) $ are given in 
Eq.(24) and Eq.(35), we obtain 
\begin{widetext}
\begin{eqnarray}
V_{LL} &=&\frac{\ \pi ^{2}\phi ^{4}}{6\ }\ [\text{ }y+\frac{3}{8}y^{2}L_{1}+%
\frac{1}{24}y^{2}L_{2}+\frac{9}{2}x^{2}L_{3}+\frac{9}{64}y^{3}L_{1}^{2}+%
\frac{1}{32}y^{3}L_{1}L_{2}+(-\frac{9}{16}xy^{2}+\frac{27}{8}%
x^{2}y)L_{1}L_{3} \\
&&+\frac{1}{576}y^{3}L_{2}^{2}+(-\frac{1}{16}xy^{2}+\frac{3}{8}%
x^{2}y)L_{2}L_{3}+(\frac{5}{16}xy^{2}-\frac{15}{8}x^{2}y+\frac{15}{4}%
x^{3})L_{3}^{2}\text{ }]  \nonumber
\end{eqnarray}
\end{widetext}
In order to obtain the total RG improved effective potential, one should incorporate the
counter-term as
\begin{equation}
V_{tot}=V_{LL}+K\phi ^{4}
\end{equation}
where $K$ can be determined by the application of the renormalization condition 

\begin{equation}
\left[ \frac{d^{4}V_{eff}(\phi )}{d\phi ^{4}}\right] _{\phi =\mu }=4\pi ^{2}y
\end{equation}
From Eqs.(40)-(42), we can obtain 
\begin{widetext}
\begin{eqnarray}
V_{tot} &=&\ \frac{\ \pi ^{2}\phi ^{4}}{6\ }\{\text{ }y+(\frac{5}{12}y^{2}+%
\frac{9}{2}x^{2})(L-\frac{25}{6})+(\frac{25}{144}y^{3}-\frac{5}{16}xy^{2}+%
\frac{15}{8}x^{2}y+\frac{15}{4}x^{3})(L^{2}-\frac{35}{3}) \\
&&+[(\frac{5}{16}y^{3}-\frac{9}{16}xy^{2}+\frac{27}{8}x^{2}y)\delta _{1}+(%
\frac{5}{144}y^{3}-\frac{1}{16}xy^{2}+\frac{3}{8}x^{2}y)\delta _{2}+\frac{15%
}{2}x^{3}\delta _{3}](L-\frac{25}{6})\}  \nonumber
\end{eqnarray}
\end{widetext}
where

\begin{equation}
\delta _{i}\equiv L_{i}-L
\end{equation}
The $\delta _{i}$ dependent terms corresponds to the difference between the single 
mass scale and the multi-mass scales $L_{i}$. Then we can obtain the ratio of the
scalar field and gauge field as 

\begin{equation}
\frac{m_{\phi }^{2}}{m_{A}^{2}} =\left[ \frac{V_{tot}^{\prime \prime
}( \phi )}{e^{2} \phi^{2}} \right]_{\phi = \mu}=r_{s} + \delta r
\end{equation}
where $r_{s}$ contains those terms coming from single mass scale case such as

\begin{widetext}
\begin{equation}
r_{s}=\frac{y}{2x}-\frac{1}{4x}(\frac{5}{2}y^{2}+27x^{2})-\frac{1}{4x}(\frac{%
275}{72}y^{3}-\frac{55}{8}xy^{2}+\frac{165}{4}x^{2}y+\frac{165}{2}x^{3})
\end{equation}
\end{widetext}
and $\delta r$ contains those terms coming from multi-mass scale such as

\begin{widetext}
\begin{equation}
\delta r=-\frac{1}{4x}[(\frac{15}{8}y^{3}-\frac{27}{8}xy^{2}+\frac{81}{4}%
x^{2}y)\delta _{1}+(\frac{5}{24}y^{3}-\frac{3}{8}xy^{2}+\frac{9}{4}%
x^{2}y)\delta _{2}+45x^{3}\delta _{3}]  
\end{equation}
\end{widetext}
The relation between $x$ and $y$ can be determined from the condition $%
\left[ dV_{tot}/d\phi \right]_{\phi = \mu} =0$ for single mass scale case and there exist
two different values of the coupling constants $y$ for each given values of $x$ \cite{Pert}. 
In table I, we give  $r_{s}$ , $\delta r$ and their ratio for typical values of the coupling constants $x$ and $y$ . As we can see in this table, their ratio becomes 
important in case of large value of $y$ (strong $\lambda$ phase ) for given $x$.

\begin{table}
\caption{\label{1} The difference between the prediction of $ \frac{m_{\phi }^{2}}{m_{A}^{2}} $ 
in case of single mass scale case ($r_{s}$) and that of multi-mass scale case ($ \delta r $) and 
absoulte value of their ratio $ r $ for the typical values parameters x and y }
\begin{ruledtabular}
\begin{tabular}{ccccc}
$x$ & $y$ & $r_{s}$ & $\delta r$ & $r=\left| \frac{\delta r}{r_{s}}\right| $ \\ \hline
0.01 & 0.0017 & 0.00151 & 0.00137 & 0.09095 \\ 
0.01 & 0.394 & 4.31415 & -5.94373 & 1.37773 \\ 
0.03 & 0.0168 & 0.04819 & 0.00151 & 0.03126 \\ 
0.03 & 0.384 & 1.43934 & -1.74417 & 1.21197 \\ 
0.05 & 0.0547 & 0.09435 & -0.01865 & 0.19769 \\ 
0.05 & 0.363 & 0.71967 & -0.90571 & 1.25850 \\ 
0.07 & 0.1536 & 0.19317 & -0.12132 & 0.62804 \\ 
0.07 & 0.2757 & 0.36282 & -0.36215 & 0.99816
\end{tabular}
\end{ruledtabular}
\end{table}

\section{Discussions and Conclusions}

In this paper, we have obtained the RG improvement of the leading-logarithms of
the effective potential for the CW model by using the method of RG
improvement in case of the multi-mass scale. Then we have investigated the
effect of the multi-mass scale on the prediction of the Higgs boson mass in case of the
two-loop order effective potential where the difference between the case of single mass
scale and that of multi-mass scale appears for the first time. We have seen that the
effect of the multi-mass scale dependent terms on the prediction of the Higgs mass 
becomes important in case of strong $\lambda$ phase. This fact implies
that when we expand the CW model to the more realistic model, one should take 
account of the multi-mass scale of the effective potential.

ACKNOWLEDGMENTS

This research was supported by the Institute of the Basic Science Research Center.

\end{document}